\documentclass[journal]{IEEEtai}

\usepackage[colorlinks,urlcolor=blue,linkcolor=blue,citecolor=blue]{hyperref}

\usepackage{color,array}

\usepackage{graphicx}
\usepackage{multirow}
\usepackage{amsmath}

\usepackage{booktabs}

\newcommand{\convMatrix}[1]{
    \begin{math}
        \left.
            \begin{array}{c}
                Conv2D + BN \\
                Conv2D + BN \\
                Conv2D + BN
            \end{array} 
        \right] \times #1
    \end{math}
}
\newcommand{\actvMatrix}[0]{
    \begin{math}
            \begin{array}{c}
                \text{ReLU} \\
                \text{ReLU} \\
                \text{ReLU}
            \end{array} 
    \end{math}
}

\newcommand{\convRow}[1]{
    \begin{math}
        \begin{array}{c}
            #1 
        \end{array} 
    \end{math}

}

\newcommand{\multiline}[2]{
    \begin{math}
            \begin{array}{c}
                #1 \\
                #2
            \end{array} 
    \end{math}
}

\begin{document}

\title{Audio Representation Learning by Distilling Video as Privileged Information}

\author{Amirhossein Hajavi, \IEEEmembership{Student Member, IEEE}, Ali Etemad, \IEEEmembership{Senior Member, IEEE}
\thanks{The authors would like to thank IMRSV Data Labs for their support of this work. The authors would also like to acknowledge the Natural Sciences and Engineering Research Council of Canada (NSERC) for supporting this research (grant number: CRDPJ 533919-18).}
\thanks{The Authors of this work are in Electrical and Computer Engineering Department of the Queen's University at Kingston, Canada. (e-mail: \{a.hajavi, ali.etemad\}@queensu.ca).}

}

\markboth{IEEE Transactions on Artificial Intelligence}{IEEE Transactions on Artificial Intelligence}

\maketitle

\begin{abstract}
Deep audio representation learning using multi-modal audio-visual data often leads to a better performance compared to uni-modal approaches. However, in real-world scenarios both modalities are not always available at the time of inference, leading to performance degradation by models trained for multi-modal inference. In this work, we propose a novel approach for deep {audio} representation learning using audio-visual data when the video modality is absent at inference. For this purpose, we adopt teacher-student knowledge distillation under the framework of learning using privileged information (LUPI). While the previous methods proposed for LUPI use soft-labels generated by the teacher, in our proposed method we use \textit{embeddings} learned by the teacher to train the student network. We integrate our method in two different settings: sequential data where the features are divided into multiple segments throughout time, and non-sequential data where the entire features are treated as one whole segment. In the non-sequential setting both the teacher and student networks are comprised of an encoder component and a task header. We use the embeddings produced by the encoder component of the teacher to train the encoder of the student, while the task header of the student is trained using ground-truth labels. In the sequential setting, the networks have an additional aggregation component that is placed between the encoder and task header. We use two sets of embeddings produced by the encoder and aggregation component of the teacher to train the student. Similar to the non-sequential setting, the task header of the student network is trained using ground-truth labels. We test our framework on two different audio-visual tasks, namely speaker recognition and speech emotion recognition. Through these experiments we show that by treating the video modality as privileged information for the main goal of audio representation learning, our method results in considerable improvements over sole audio-based recognition as well as prior works that use LUPI.
\end{abstract}


~\\
\section{Introduction}

\IEEEPARstart{D}{eep} {audio} representation learning has recently attracted significant interest, specially in applications such as speaker recognition (SR) \cite{CNN_speech_mine_short, CNN_speech_mine, xie2019utterance, nagrani2020voxceleb, hajavi2021siamese} and speech emotion recognition (SER) \cite{albanie2018emotion, jalal2019learning, kumar2021towards}. The goal in deep {audio} representation learning is to learn embeddings from audio or visual signals, which could be used in retrieving information such as identity or the emotional state of the speaker. This goal is generally best achieved when \textit{multi-modal} audio-visual inputs are used \cite{nagrani2020voxceleb, kansizoglou2019active, atmaja2020multitask} as opposed to when only a single modality of audio or video is used \cite{CNN_speech_mine_short, CNN_speech_mine, xie2019utterance, tedd1, tedd2, hajavi2021siamese}. Nonetheless, in many real-world scenarios, both modalities may not be simultaneously available at \textit{inference}, resulting in the inability of the model to perform effectively. To tackle this, we pose the question: ``\textit{how can training with both modalities be performed effectively to benefit inference with a single modality?}''



\begin{figure}[!t]
  \centering
  \includegraphics[width=.7\columnwidth]{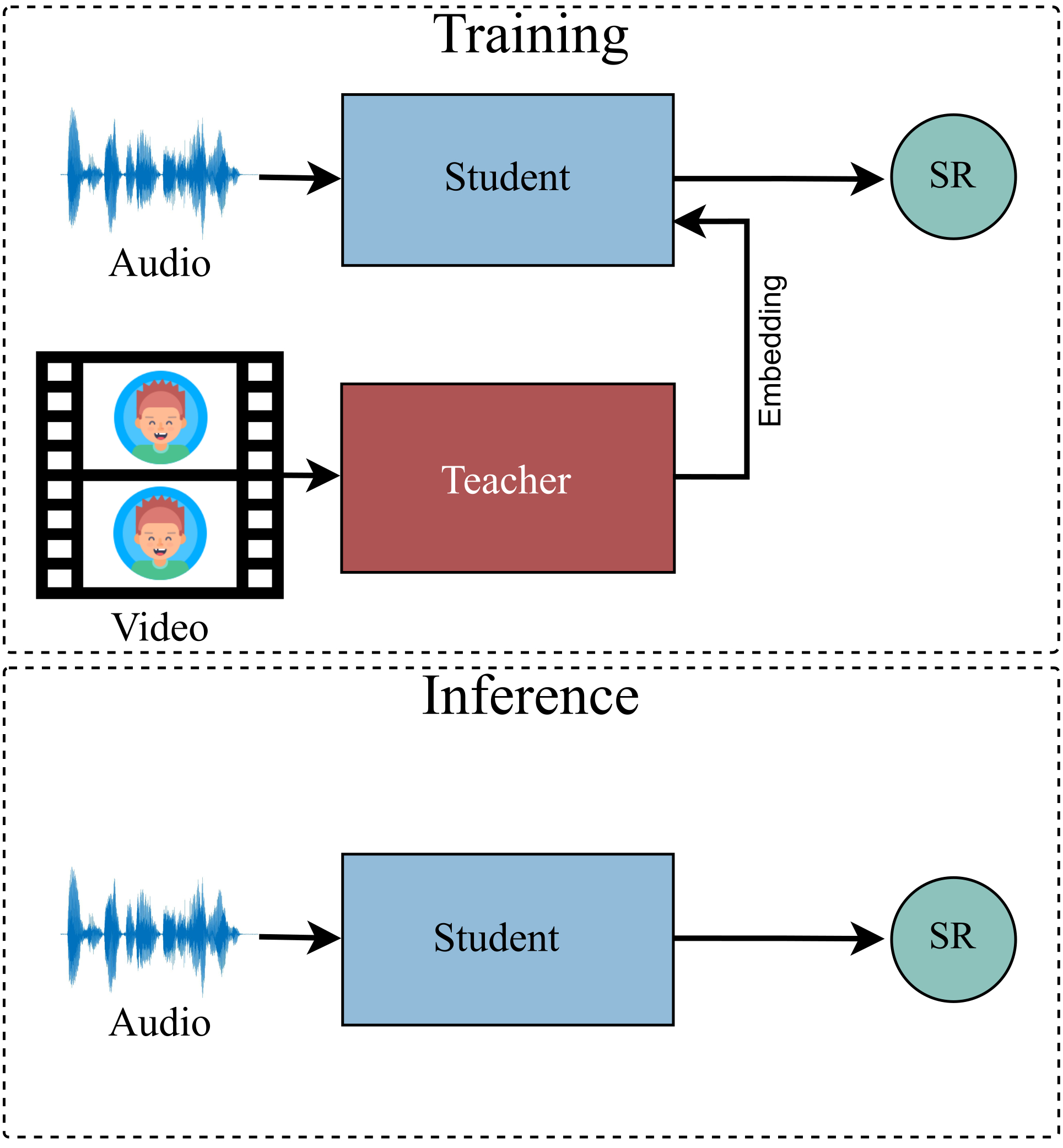}
  \caption{Overview of the proposed method. {Embeddings extracted from the video modality by the teacher are used as privileged information in training the student to boost its ability in learning audio representations. At inference, only the audio modality is present and the student model is tasked with generating audio embeddings which are then used for performing SR and SER.}}
  \label{fig:model_overview}
\end{figure}

The study done by Vapnik et al. \cite{vapnik2009LUPI1} defined information only available during training (and not at inference) as ``privileged information''. They introduced a new learning paradigm called `learning using privileged information' (LUPI) in which a secondary SVM model trained on the privileged information helped the main SVM to perform better on its task by reducing the complexity of the problem by optimizing the slack variables. This paradigm was later adapted into deep neural networks in \cite{vapnik2016LUPIMAIN}, showing that LUPI can be performed using knowledge distillation techniques proposed by Hinton et al. in \cite{hinton2015distilling}. In their work, a teacher model was trained using the privileged information using a textual modality. The teacher along with the ground-truth labels were then used to train the student model to perform image classification. 

In this study, to perform uni-modal {audio representation learning} while training with multi-modal audio-visual streams, we propose a novel solution by adopting privileged information and considering video as such. An overview of our method is shown in \mbox{Figure \ref{fig:model_overview}}. Our model is built based on teacher-student knowledge distillation to allow for the video stream to be learned alongside the main audio modality during training. First, we train a teacher network using the video stream as the privileged information. Next, our student network is trained using audio as input and the ground-truth output labels as well as the video embeddings obtained from the teacher, simultaneously as outputs. By doing so, our student model can perform solely on the audio signals during inference, while having been trained with and benefited from both audio and video modalities during training. We perform extensive experiments using multiple widely used audio-visual datasets to demonstrate the effectiveness of the proposed framework in exploiting the privileged information for {audio representation learning}.

In summary, we make the following contributions:
\begin{itemize} 
    \item We propose a new solution for deep {audio representation learning} for SR and SER that utilizes video as privileged information during the training of the network to improve its performance.
    \item We perform SR and SER experiments on our model and observe considerable improvements versus uni-modal baselines. We also compare our approach to other studies based on privileged information and show that our method performs better for {audio representation learning}.
    \item We provide an analysis on the impact of the privilege information by adjusting its influence on training the networks. 
\end{itemize}


The rest of the paper is organized as follows. {Section \ref{sec:related_work} presents the previous work on privileged information and knowledge distillation. In section \ref{sec:method}, we present a detailed description of our proposed solution. Afterwards, we describe the performed experiments in detail and report the results in section \ref{section:experiments}. Finally we conclude the work with a summary and discussion on potential future directions. }

\section{Related Work}
\label{sec:related_work}
Different approaches have been taken in the literature for learning with the help of privileged information \cite{albanie2018emotion, do2019compact, gao2021residual, garcia2018modality, passalis2018learning, roheda2018cross, tian2019contrastive, Nagrani_2018_ECCV, thoker2019cross,  nagrani2020disentangled, Nawaz_2021_CVPR, shi2017learning, lambert2018deep}. We can categorize these studies into two main groups: (1) Those that rely on knowledge distillation techniques via teacher-student models for deep representation learning; (2) Those that utilize privileged information without the use of knowledge distillation. In the following, we first review the general concept of knowledge distillation given its relevance in LUPI as well as our method. This is followed by a review of LUPI with and without adopting knowledge distillation. While our work lies in the field of {audio} representation learning, given the low number of works in the area of using privileged information for audio, we expand our discussion to other modalities to provide a more comprehensive review.

\subsection{Knowledge Distillation}
Knowledge distillation was proposed by Hinton et al. \cite{hinton2015distilling} to enable smaller machine learning models (referred to as `student') to learn from larger machine learning models (referred to as `teacher'). A large number of studies have since explored the use of knowledge distillation for deep representation learning. Seminal works in this area include \cite{romero2015fitnets, huang2017like, kim2018paraphrasing, zhou2018rocket, heo2019comprehensive, xu2020feature, guan2020differentiable, heo2019knowledge}. These studies demonstrated that the use of knowledge distillation enables student models to achieve competitive performances to the teachers, while reducing the number of learnable parameters, hence computational load and required memory. 

Knowledge distillation in general is performed via two main approaches. The first approach is to use the `\textit{soft labels}' obtained from the teacher to train the student \cite{you2017learning, zhang2018better}, while the second approach instead relies on the `\textit{embeddings}' learned by the teacher to train the student \cite{romero2015fitnets, huang2017like, kim2018paraphrasing, zhou2018rocket, heo2019comprehensive, xu2020feature, guan2020differentiable}. In addition to these two approaches, a few other solutions have also been proposed in the literature. For instance, the activation boundaries of the neurons in the teacher were transferred to the student in \cite{heo2019knowledge} to reduce the training time. In \cite{komodakis2017paying, passban2021alp}, knowledge distillation was performed by transferring the attention maps generated by the teacher to the student. This technique helped the student to find the salient areas of the input with the help of the transferred attention maps, which in turn boosted the performance of the student without the need for additional training. In \cite{chen2021cross}, instead of relying only on a single embedding, the knowledge from multiple layers of the teacher was used to train the student, boosting the generalizability of the student during inference.

\subsection{LUPI with Knowledge Distillation}

As one of the earliest solutions for LUPI with deep neural networks, the use of teacher-student knowledge distillation was proposed in \cite{vapnik2016LUPIMAIN}. 
In their work, it was proposed that the teacher can receive its input from the privileged information and its output can be used alongside the ground-truth labels to train the student which receives its input from the main data. 
A number of recent literature have used similar techniques in their studies \cite{albanie2018emotion, do2019compact, gao2021residual, garcia2018modality, passalis2018learning, roheda2018cross, tian2019contrastive, Nagrani_2018_ECCV, thoker2019cross,  nagrani2020disentangled, Nawaz_2021_CVPR}. The main idea behind these methods has been to use a secondary modality for the input of the teacher model as the source of privileged information. 

A very limited number of prior works have targeted {audio} representation learning while considering `\textit{video}' as the secondary modality \cite{albanie2018emotion, nagrani2020disentangled, Nawaz_2021_CVPR}. In their solutions, the teacher model takes video frames as input and generates soft-labels which are then used as the only source for training the student. The main aim of these studies is to alleviate the need for labeled training data in the main modality (audio) by training the student models using only the soft-labels generated from a secondary modality (video). While these studies successfully achieve their goal of training audio using video as a [self-] supervisory signal, they do not explore the impact of using the secondary modality as privileged information for helping the networks in learning the main modality. However, in this work we aim to take advantage of the secondary modality to boost the performance of the networks alongside the labeled data from the main modality instead of avoiding the use of output labels altogether. 


\subsection{LUPI \textit{without} Knowledge Distillation}
\label{sec:relate_lupi_wo}
A number of studies have also taken approaches other than teacher-student frameworks toward utilizing privileged information for training their networks. In the study performed by \cite{shi2017learning}, multi-task learning was used for training the model. Their proposed network performs action recognition while also aiming to reconstruct the privileged information from a different modality. The main modality used in their study was the video frames of individuals performing specific actions while for the privileged information, the positions of skeletal joints of individuals in the videos were used. The use of privileged information in this work boosted the performance compared to several baselines. While their proposed method proved beneficial for secondary modalities with limited dimensionality, its integration with secondary modalities with high dimensionality, for instance video frames, has not been explored.
In \cite{lambert2018deep}, dropout masks derived from privileged information were used in order to generalize a DNN. The study used privileged information in the form of image segments obtained from the input to generate heuristic dropout masks. These masks were then applied over the learnt representations of the DNN in order to help the model with generalization. While the method was shown to boost the generalizability of models, the dropout masks are generated using a segment of the original input. Hence this method requires the privileged information to be sourced from the same modality as the original input.

\section{Method}
\label{sec:method}
Our objective in this study is to train a deep neural network to learn {audio} representations using audio-visual data under the condition that the video modality is not available during inference/testing. Our approach, adopts the paradigm of teacher-student knowledge distillation and LUPI to train a network that can handle this condition. In our approach, the student model operates on the audio modality and is trained using two outputs, one from the ground-truth labels and the other from the embeddings obtained by the teacher model which operates on the secondary modality. In this section we describe the different components of our proposed method. 

\subsection{Preliminaries}
\label{sec:method_softlabels}
The training data used for supervised learning often comes in the form of tuples $(x_i, y_i)^{i\in\{0,..,n\}}$ where $x_i$ is a vector representation of the training sample and $y_i$ is the output label. The aim of the model is to find an optimal network $\mathcal{F}$, that predicts the labels of the test data $y_i = \mathcal{F}(x_i)$ with the least amount of error. In this type of training the format of data stays the same during training, testing, and deployment of the model. Occasionally, during the training phase we may have access to additional information other than what is available in the test set. Vapnik et al. referred to this kind of information as privileged information and proposed a technique called ``LUPI'' to take advantage of this information for better training machine learning models \cite{vapnik2009LUPI1}. The \textit{training} of the models through such a paradigm is done using the tuple $(x_i, x_i^*, y_i)$ where $x_i^*$ represents the privileged information. However, during \textit{inference} with the trained models, the data would still maintain its previous format and the additional information is not available. 

The LUPI paradigm was first introduced in the context of support vector machine (SVM) \cite{vapnik2009LUPI1}. The model, namely SVM+, was shown to perform better than the classical SVM, i.e. SVM trained without privileged information.
The LUPI paradigm can also be implemented using teacher-student knowledge distillation. In this view of the paradigm, the original knowledge distillation method \cite{hinton2015distilling} is expanded by using the privileged information as the input for the teacher model. Given the tuple $(x_i, x^*_i, y_i)$, the teacher model is trained using the tuples of privileged information and label $(x^*_i, y_i)$ at first. In the next step, soft-labels $s_i$ are obtained for each training sample from the teacher as the predicted output probability, using the privileged information. Finally,  {any layer $L$ of the} student model is trained using the tuples $(x_i, y_i)$ and $(x_i, s_i)$ concurrently with the final gradient $\nabla_s$ calculated as follows:
\begin{equation}
    \nabla_s = (1-\alpha) \nabla \frac{\mathcal{L}_i(y'_i, y_i)}{\nabla L} + \alpha \nabla \frac{\mathcal{L}_i(y'_i, s_i)}{\nabla L}.
    \label{eq:normal_lupi}
\end{equation}
Here the parameter $\alpha$ is the imitation parameter that determines how much the student model should follow the teacher. 

\subsection{Our Solution}
As described earlier, our goal is to develop a framework capable of distilling video data into an audio learner, so that training is improved given the availability of both modalities, while audio alone is used at inference. For this purpose we use teacher-student knowledge distillation in which the teacher network operates on video data, while the embeddings extracted from its intermediate layers are used to help train the student network which operates on audio data. An overview of our method was depicted earlier in \mbox{Figure \ref{fig:model_overview}}. Through the following subsection the details of each component in our approach is presented. It should be noted that similar to most works on audio representation learning, we use spectrograms extracted from the audio signals in our model. However, these spectrograms can be considered in two different formats: non-sequential and sequential. The non-sequential representation of the audio considers the entire spectrogram as one single image, while in the sequential version, the spectrograms are divided into smaller segments across time. Accordingly, we design our model for both approaches (non-sequential and sequential) and present our teacher-student method for both versions in the following subsections.

\subsubsection{Video Learning}

Here we describe the process of video representation learning in our model, which is carried out by the teacher network. {The process of training the teacher network is done separately and happens prior to the training of the student network. During the training of the student, the teacher network is frozen and its weights are unchanged.} Let us assume the training samples are represented as a set of tuples $(x_i, x^*_i, y_i)$, where $x_i$ is the $i^{th}$ training sample from the audio modality and $x^*_i$ is the corresponding sample from the video modality. In the non-sequential approach, the teacher network is comprised of an encoder, $F^{T_{enc}}$, accompanied by the task header which includes a Fully Connected (FC) layer followed by softmax activation, denoted by $F^{T_{head}}$. Here $F^{T_{enc}}$ generates embeddings for each frame of the video from $x^*$ using 
\begin{equation}
    E^T_{i_j} = F^{T_{enc}}({x^*_{i_j}}),
    \label{fml:encoder_embedding}
\end{equation}
where $j$ is the video frame index. The accompanying $F^{T_{head}}$ then generates the predicted labels of the model. The final output of the teacher model $O^T$ is accordingly calculated as 
\begin{equation}
    O^T_{i_j} = F^{T_{head}}(E^T_{i_j}). 
\end{equation}
The final output $O^T_{i_j}$ comes in the form of a vector with a dimensionality equal to the number of classes considered in by the model. Each index of $O^T_{i_j}$ contains the calculated probability of $x^*_{i_j}$ belonging to the class represented by the index. 

In general, when learning audio-video representations, several video frames often correspond to a spectrogram spanning over a period of time. To address this, prior works have proposed the selection of a single frame known as \textit{peak frame} \cite{albanie2018emotion,nagrani2020disentangled, Nawaz_2021_CVPR} for each given input spectrogram. Inspired by this approach, we select a peak frame embedding in our solution as per the following:
\begin{equation}
      E^T_{i_{peak}} = E^T_{i_j} | Arg\!\max_{j}(Arg\!\max(O^T_{i_j})),
\end{equation}
where the frame with the highest calculated probability for its class is selected. {The class is determined by performing an $Arg\!\max$ on the $O^T_{i_j}$ by the softmax function}.

On the other hand, in the sequential approach, we use the entire video as the input for the teacher network. In this version, an aggregation sub-network, $F^T_{agg}$,  is added to the teacher network and is placed between the encoder and the task header. Here $F^{T_{agg}}$ generates an embedding from the output of the encoder using:
\begin{equation}
      E^{T_{agg}}_{i} = F^{T_{agg}}(\{E^T_{i_j}| 0\le j < N_i\}) ,
      \label{fml:lstm_embedding}
\end{equation}
where $E^T_{i_j}$ are the encoder embeddings obtained by Eq.\ref{fml:encoder_embedding} and $N_i$ is the number of frames in each video. 

When taking the sequential approach, we also divide the input $x_i$ of the audio into fixed sized segments $x_{i_k}$ where $k$ is the index of the audio segment. As mentioned earlier, spectrograms that span across a period of time often cover multiple frames of video. Therefore, each segment of the audio $x_{i_k}$ will be matched with multiple embeddings $E^T_{i_j}$. In the sequential version of the teacher network, each frame does not have a probability score for its class and therefore selection of a peak frame is not possible. To address this issue we define the embedding $E^T_{i_k}$ {as the average, $Avg$, of all of the embeddings in the segment}:
\begin{equation}
      E^T_{i_k} = Avg(\{E^T_{i_j}|r\times k\le j < r\times (k+1)\}),
      \label{fml:encoder_embedding_segmented}
\end{equation}
where $r$ is the number of frames per audio segment. The calculated embedding $E^T_{i_k}$ is then matched with the audio segment $x_{i_k}$.

\begin{figure}[!t]
  \centering
  \includegraphics[width=0.8\columnwidth]{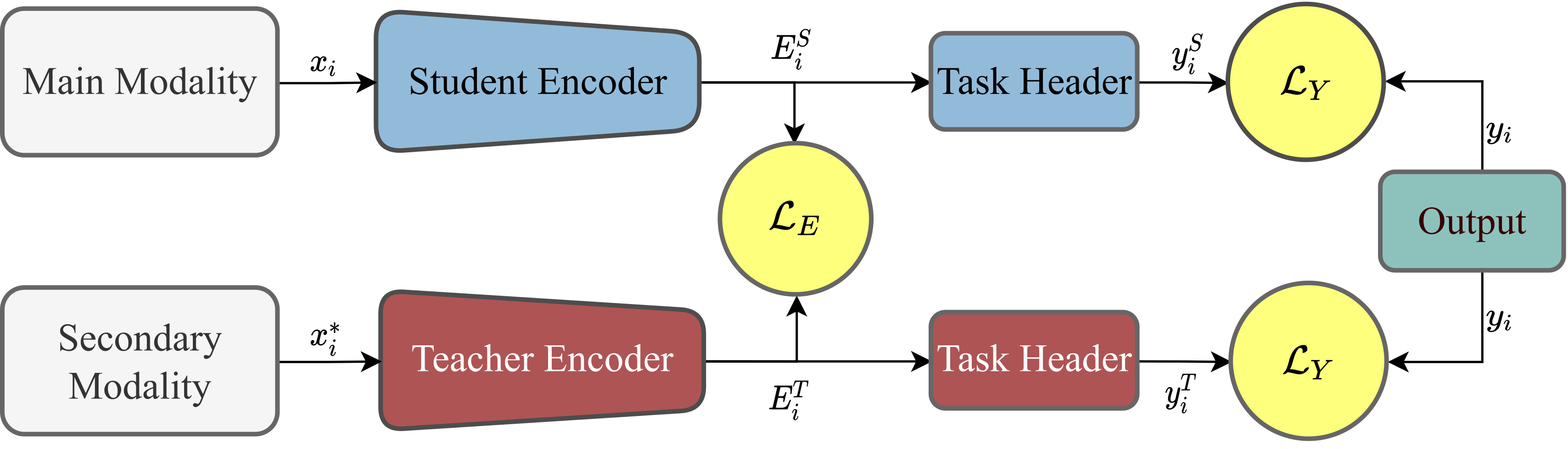}
  \caption{The proposed framework for learning privileged information through teacher-student distillation, in non-sequential settings. {In this setting the entirety of the audio signal is given to the student as a single spectrogram and the teacher network receives a single frame of the video as input. Here, the student and teacher networks are both comprised of an encoder component and a task header. The embeddings generated by the encoder component of the teacher network are used in partially training the encoder component of the student network using gradients generated by $\mathcal{L}_E$. This is while the encoder component and the task header of the student network receive gradients generated from $\mathcal{L}_Y$}. }
  \label{fig:model_non}
\end{figure}

\begin{figure*}[!t]
  \centering
  \includegraphics[width=1\textwidth]{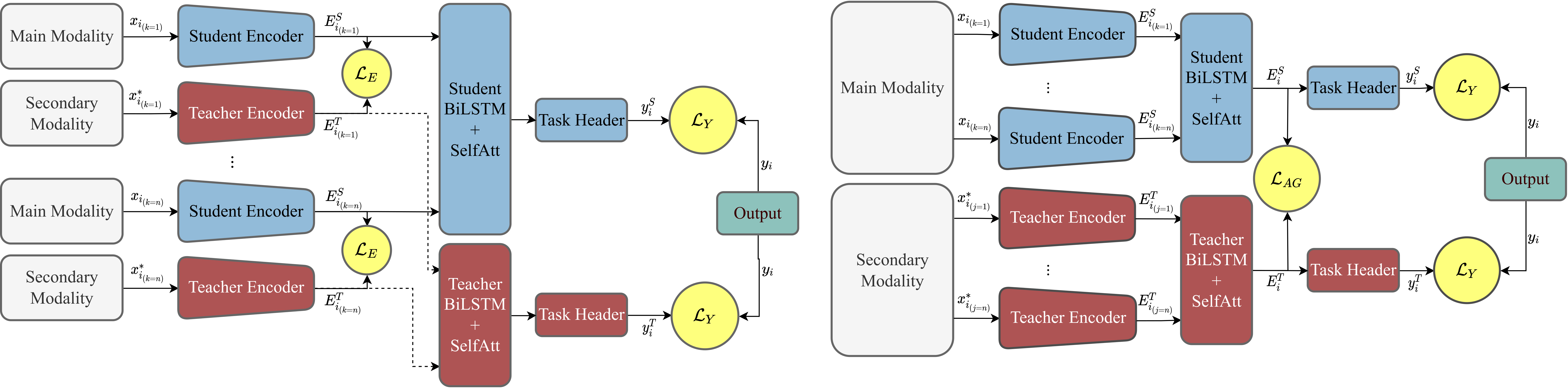}
  \caption{The proposed framework for learning privileged information through teacher-student distillation.{ In the sequential setting of our method, the audio signal is divided into multiple same-sized sections throughout time. Each section is then given to the encoder component of the student network. The embeddings generated by the encoder component are then collected and passed onto an aggregator which extracts time dependencies from these embeddings. The teacher network on the other hand generates embedding for each video frame as well as an overall embedding of the entire video. In the encoder-level implementation of our method (left), part of the gradients used for training the encoder of the student network are obtained from $\mathcal{L}_E$, which compares the embeddings generated by the encoder of the student network and embeddings generated by the encoder of the teacher network. In aggregator-level implementation of the proposed method (right), both encoder and aggregator components of the student network receive gradients from privileged information through $\mathcal{L}_{AG}$. This loss compares the embeddings generated by the aggregator of the student network with the embeddings generated by the aggregator of the teacher network. In both implementations, the other part of the gradients which are also used for the remainder of the student network are generated using ground truth labels.}}
  \label{fig:model_seq}
\end{figure*}

\subsubsection{Audio Learning}

In the non-sequential approach the student network takes in the entirety of the audio signal in a single spectrogram. \mbox{Figure \ref{fig:model_non}} shows the general scheme for the non-sequential version of the proposed method. In this version, the student model is comprised of an encoder, $F^{S_{enc}}$, and a task header which consists of an FC layer with a softmax activation, denoted by $F^{S_{head}}$. The encoder extracts embeddings $E^S_i$ from the input $x_i$ using 
\begin{equation}
    E^S_i = F^{S_{enc}}(x_i).
\end{equation}
The embedding $E^S_i$ is then given to $F^{S_{head}}$ to predict the output $y'_i$. We then define a loss function $\mathcal{L}_E(E^S_i, E^T_{i_{peak}})$, which calculates the distance between the embeddings of the encoder component of the student network and the embedding calculated using Eq.\ref{fml:encoder_embedding}. We define a second loss function $\mathcal{L}_Y(y'_i, y_i)$, which calculates the distance between $y'_i$ and the ground-truth labels $y_i$. The output layer of the student model is trained using only $\mathcal{L}_Y$, whereas the encoder is trained by the gradient $\nabla_s$ calculated using
\begin{equation}
    \nabla_s = (1-\alpha) \nabla \frac{\mathcal{L}_Y(y'_i, y_i)}{\nabla F^{S_{enc}}} + \alpha \nabla \frac{\mathcal{L}_E(E^S_i, E^T_{i_{peak}} )}{\nabla F^{S_{enc}}},
\end{equation}
where $\alpha$ acts as the imitation parameter. The value of $\alpha$ defines the weight of the gradients while training the encoder component of the student and ranges from $0$ to $1$. 

In the sequential approach, the audio input of the student model is divided into fixed-sized segments across time. Here, an aggregator sub-network $F^{S_{agg}}$ is added to the student network between the encoder and the task header. The encoder $F^{S_{enc}}$ generates embeddings $E^S_{i_k}$ from each audio segment $x_{i_k}$ using
\begin{equation}
    E^S_{i_k} = F^{S_{enc}}(x_{i_k}).
\end{equation}
The final embedding $E^{T_{agg}}_{i}$ is then calculated using 
\begin{equation}
      E^{T_{agg}}_{i} = F^{S_{agg}}(\{E^T_{i_k}| 0\le k < M_i\}),
\end{equation}
where $M_i$ is number of audio segments in $x_i$. Lastly, the output of the student model $y'_i$ is generated using
\begin{equation}
    y'_i = F^{S_{head}}(E^{T_{agg}}_{i}).
\end{equation}

\begin{table}[!t]
    \centering
    \footnotesize
    \caption{Architecture details for the VGG-based non-sequential student networks.}
    \begin{tabular}{l|c|l}
        \hline
        Layer & Activation & Shape \\
        \hline
        \hline
        \convRow{Input}               & --       &  $257 \times  500  \times 1$      \\\hline
        \convRow{Conv2D+BN}           & ReLU     &  $257 \times  250  \times 64$     \\
        \convRow{Maxpool}             & --       &  $128 \times  250  \times 64$     \\\hline
        \convRow{Conv2D+BN}           & ReLU     &  $128 \times  250  \times 128$    \\ 
        \convRow{Maxpool}             & --       &  $64  \times  125   \times 128$   \\ \hline
        \convRow{Conv2D+BN}           & ReLU     &  $64  \times  125   \times 128$   \\ 
        \convRow{Maxpool}             & --       &  $32  \times   62   \times 128$   \\\hline
        \convRow{Conv2D+BN}           & --       &  $32  \times   62   \times 256$   \\
        \convRow{Conv2D}              & ReLU     &  $32  \times   62   \times 256$   \\ 
        \convRow{Maxpool}             & --       &  $16  \times   31   \times 256$   \\\hline
        \convRow{Conv2D+BN}           & ReLU     &  $16  \times   31   \times 512$   \\ 
        \convRow{Maxpool}             & --       &  $8   \times   15    \times 512$  \\ \hline
        \convRow{Conv2D+BN}           & ReLU     &  $8   \times   15    \times 512$  \\ 
        \convRow{Maxpool}             & --       &  $4   \times   15    \times 512$  \\\hline
        \convRow{FC}                  & --       &  $512$                            \\ 
        \convRow{FC}                  & ReLU     &  $512$                            \\
        \convRow{Output}              & Softmax  &  \# of Classes              \\ 
        \hline
    \end{tabular}
    \label{tab:vgg_non_seq}
\end{table}

\begin{table}[!t]
    \centering
    \footnotesize
    \caption{Architecture details for the ResNet and SEResNet -based non-sequential student networks.}
    \begin{tabular}{l|c|l}
        \hline
        Layer & Activation & Shape \\ \hline\hline
        \convRow{Input}             &     --                &  $257 \times  500  \times 1$  \\ \hline
        \convRow{Conv2D}            &     ReLU              & $ 257 \times 500 \times  64 $ \\
        \convRow{Maxpool}           &     --                & $ 128 \times 250 \times  64 $ \\ \hline
        \convMatrix{2}              &     \actvMatrix       & $ 128 \times 250 \times  96 $ \\ \hline   
        \convRow{Maxpool}           &     --                & $  64 \times 125 \times  96 $ \\\hline
        \convMatrix{3}              &     \actvMatrix       & $  64 \times 125 \times 128 $ \\ \hline
        \convRow{Maxpool}           &     --                & $  32 \times 62  \times 128 $ \\\hline
        \convMatrix{3}              &     \actvMatrix       & $  32 \times 62  \times 256 $ \\ \hline
        \convRow{Maxpool}           &     --                & $  16 \times 31  \times 256 $ \\\hline
        \convMatrix{3}              &     \actvMatrix       & $  16 \times 31  \times 512 $ \\ \hline
        \convRow{Maxpool}           &     --                & $  8 \times  15  \times 512 $ \\ \hline
        \convRow{FC}                &     --                &  $512$                        \\ 
        \convRow{FC}                &    ReLU               &  $512$                        \\
        \convRow{Output}            &    Softmax            &  \# of Classes                \\ \hline
    \end{tabular}
    \label{tab:resnet_non_seq}
\end{table}

\begin{table}[!t]
    \centering
    \footnotesize
    \caption{Architecture details for sequential student networks. $N$ is the number of segments in which the input is divided into.}
    \begin{tabular}{l|c|l}
        \hline
        Layer & Activation & Shape \\ \hline\hline
        \convRow{Input}             &     --       & $N \times 257 \times  100  \times 1$   \\ \hline
        \convRow{Encoder}           &     --       & $N \times 512 $                        \\ \hline
        \convRow{BiLSTM}            &     ReLU     & $N \times 512 $                        \\ 
        \convRow{BiLSTM}            &     ReLU     & $N \times 512 $                        \\ 
        \convRow{Self-Attention}    &    Attention & $512$                        \\ \hline
        \convRow{Output}            &    Softmax   & \# of Classes                \\ \hline
    \end{tabular}
    \label{tab:seq_model}
\end{table}

\mbox{Figure \ref{fig:model_seq}} shows the two possible approaches for the sequential version of the proposed method for training the student networks: (1) distilling video information at the encoder-level shown in \mbox{Figure \ref{fig:model_seq}} (left); (2) distilling video information at the aggregator-level shown in \mbox{Figure \ref{fig:model_seq}} (right). 
In the encoder-level distillation we use the loss function $\mathcal{L}_{E}(E^S_{i_k}, E^T_{i_k})$ which calculates the loss between the embedding of each audio segment $E^S_{i_k}$ and the teacher embedding $E^T_{i_k}$ corresponding to that segment. 
The encoder component of the student model $F^{S_{enc}}$ is then trained by the gradients calculated using 
\begin{equation}
    \nabla_s = (1-\alpha) \nabla \frac{\mathcal{L}_Y(y'_i, y_i)}{\nabla F^{S_{enc}}} + \alpha \sum_{k=0}^{M_i} \nabla\frac{\mathcal{L}_E(E^S_{i_k}, E^T_{i_k} )}{\nabla F^S},
\end{equation}
while the aggregator and the output layer of the student network are trained only using gradients calculated by the loss function $\mathcal{L}_Y(y'_i, y_i)$.
 
In the aggregator-level distillation we use the loss function $\mathcal{L}_{AG}(E^{S_{agg}}_{i}, E^{T_{agg}}_{i})$, which calculates the distance between the embeddings generated by the aggregator component of the student model and the embedding $E^{T_{agg}}_{i}$ calculated using Eq.\ref{fml:lstm_embedding}. The output layer of the student network is train using only $\mathcal{L}_Y(y'_i, y_i)$, while the rest of the pipeline is trained by the gradients calculated using
\begin{equation}
    \nabla_s = (1-\alpha) \nabla \frac{\mathcal{L}_Y(y'_i, y_i)}{\nabla F^{S_{agg}}} + \alpha \nabla \frac{\mathcal{L}_{AG}(E^{S_{agg}}_{i}, E^{T_{agg}}_{i} )}{\nabla F^{S_{agg}}}.
\end{equation}



\subsection{Implementation Details}
The networks in the non-sequential version of our solution consist of an encoder and an output layer. We use 3 architectures based on VGG \cite{simonyan2014very}, ResNet \cite{he_deep_2016}, and Squeeze-and-Excitation (SE) Networks  \cite{hu2018squeeze} to implement both the teacher and student networks. For the first benchmark we use a standard VGG16 network for the teacher and a VGG-based network  customized for audio in the student. Table \ref{tab:vgg_non_seq} presents the details of the VGG-based student network. In this network each convolutional layer is coupled with a batch-normalization layer. We use a stride length of 1 in the convolutional and batch-normalization layers throughout the network. For the the maxpooling layers we use a filter size of (2, 2) and a stride length of 2. We add 2 FC layers with 512 neurons after the last maxpooling layer. The output of the last FC layer is used as the student embedding $E_s$, described earlier in Section \ref{sec:method}. We use Rectified Linear Unit (ReLU) as the activation function for the convolutional layers and the last FC layer. Lastly, for the task header we use an FC layer with the number of neurons equal to the number of classes in the experiments. We then use a softmax function for the this layer.

For the second benchmark we utilize a standard ResNet34 for the teacher network and a ResNet-based model customized for audio, in the student network. Details of the ResNet-based student model are presented in Table \ref{tab:resnet_non_seq}. The first convolutional layer of the student model uses a filter size of (7, 7) with a stride length of 1. This layer is followed by a maxpooling layer with a filter size of (2, 2) and a stride length of 2. Afterwards, we use \textit{residual blocks} in the student model. Each block contains 3 sets of coupled convolutional and batch-normalization layers and a shortcut connection that links the input of the block to its output. The first and last convolutional layers in each block have a filter size of (1, 1) and the second convolutional layer has a filter size of (3, 3) with a stride of 1. Each block is then repeated multiple times as shown in Table \ref{tab:resnet_non_seq}. We add a maxpooling layer with a filter size of (2, 2) and a stride of 2 after each block. The last maxpooling layer is then followed by 2 FC layers, each with 512 neurons and an FC layer in the task header with neurons equal to the number of classes. Similar to the first benchmark, we use ReLU as the activation functions for the convolutional layers and the last FC layer in the encoder, while softmax is used for the FC layer in the task header. 

\begin{figure*}[!t]
  \centering
  \includegraphics[width=.9\textwidth]{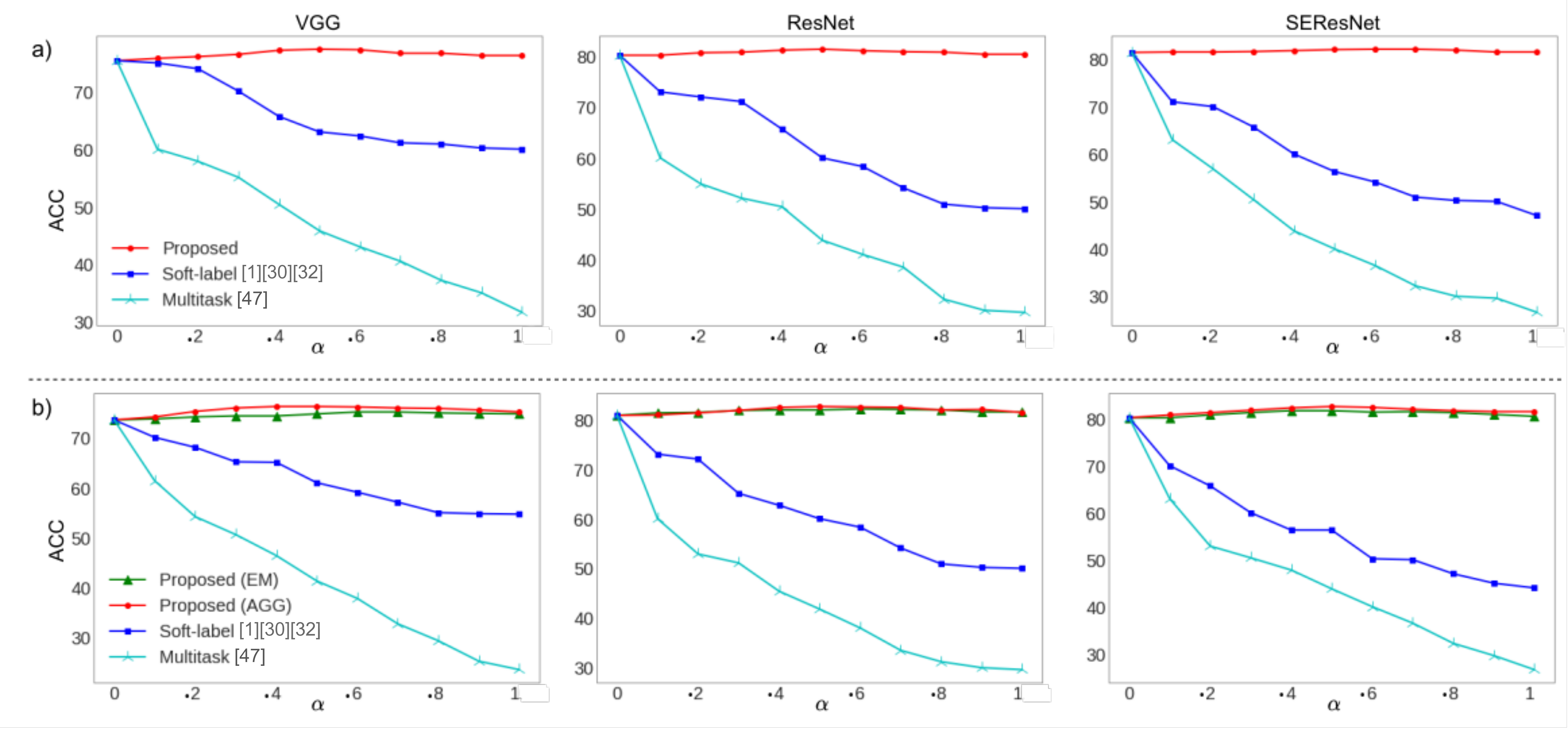}
  \caption{{A comparison between the effect of using our proposed method for LUPI versus using Soft-label and multitask training for speaker identification in relation to different values of $\alpha$ in: (a) Non-Sequential settings; (b) Sequential settings.}}
  \label{fig:alpha_speaker}
\end{figure*}

In the third benchmark we construct the networks using SE blocks. These blocks use the same layer format of the residual blocks with the difference that an SE module is added to each block. The SE module consists of a global pooling layer, which extracts channel information, 2 FC layers with a ReLU activation function in between, and a sigmoid activation function following the FC layers. We implement the teacher network by replacing the residual blocks from a standard ResNet34 network with SE blocks. The student network is implemented by replacing the residual blocks in the network from the second benchmark with SE blocks.

In the sequential version of our solution, the networks comprise of an encoder component and an aggregator component. Similar to the non-sequential format, we perform our experiments using 3 benchmark networks. For the encoder component of our teacher and student networks we use the teacher and student networks used in the non-sequential version, respectively. However, the task headers of the student networks are removed and the encoder component generates an embedding vector with a size of 512 for each segment of the input. The aggregator component of the networks include 2 BiLSTM layers accompanied by an attention module. Table \ref{tab:seq_model} shows the details of the student networks for the experiments in the sequential version of the proposed method.

\section{Experiments}
\label{section:experiments}

\subsection{Datasets and Data Preparation}
The aim of the experiments is to evaluate the change in performance between learning audio representations alone (for {audio} representation learning) versus using video as privileged information. 
The experiments are done on two different tasks of SR and SER. We use 3 publicly available datasets, namely \textbf{VoxCeleb} \cite{nagrani2020voxceleb}, Ryerson Audio-Visual Database of Emotional Speech and Song (\textbf{RAVDESS}) \cite{livingstone2018ryerson}, and \textbf{IEMOCAP} \cite{busso2008iemocap}. 

We use the audio-video version of the VoxCeleb dataset \cite{nagrani2020voxceleb} for SR. In this task we aim to identify the speaker of a given utterance among a set of known speakers. This version of the VoxCeleb dataset is comprised of 21,819 audio-visual recordings from 1,211 individuals. {We use 70\% of the recordings from all the 1,211 individuals for training, 10\% for validation, and 20\% for testing.} We use the spectrogram representations of audio as inputs to the student model. The frequency features are extracted from the audio using Short-term Fourier Transform (STFT) with a window of 25 \textit{ms}. The process is repeated across the entire utterance with a window overlap of 10 \textit{ms}. The duration of the utterances is not the same for all of the recordings. In order to rule out the complications caused by the variable length of the inputs 
all the recordings are cropped at 5-second durations {when training the models}, resulting in spectrograms of size $257\times 500$. {It should be noted that the original length of the recordings are used for inference}. The shorter utterances are padded using repetition to match the desired length. The videos are recorded at a frame-rate of 25 frames per second. Each frame is annotated using automated face detection models, giving the location and boundaries of the face of the speaking person. The size of the boundaries are not equal across the dataset, therefore we crop the images and resize them to a fixed dimension of $224\times 224$. 
For evaluation of the sequential implementation of the proposed method, the recordings are divided into 1-second segments. This results in a sequence of 5 smaller spectrograms with a dimensionality of $257\times 100$ for each utterance and 5 sets of frames for each video, with 25 frames in each set.

For SER we use two datasets, RAVDESS  \cite{livingstone2018ryerson} and IEMOCAP  \cite{busso2008iemocap}. In this task, we aim to identify the emotional state of the speaker of an utterance and classify that state into different discrete emotion categories namely \textit{Sad}, \textit{Happy}, \textit{Fearful}, \textit{Disgusted}, \textit{Surprised}, \textit{Angry}, and \textit{Neutral}. The RAVDESS dataset is comprised of recordings from 24 participants. {We use cross-validation through leave-one-subject-out for training and validating our method and report the mean accuracy}. Participants are asked to deliver a sentence with 7 different emotions in two forms of normal speech and song. Each actor performs a sentence 60 times in normal speech and 44 times in singing voice with one actor exempted from singing. The total number of video recordings used for our experiments is 2,452. The video clips are recorded under controlled conditions without any environmental noise. The length of recordings are fixed to 4 seconds, therefore no additional padding or cutting is performed on the input. The frame rate of the recorded videos is set to 25 frames per second. In most of the frames, the face of the speaker is located in the center of the frame and covers at most 50\% of the frame. Therefore, we crop the frames in the center and resize the resulting image to the fixed dimensions of $224\times 224$.

Lastly, we also use IEMOCAP for evaluating our method on SER. This dataset contains a total of 6 thousand audio-visual recordings performed by 10 individuals. {We use 5-fold cross-validation for evaluation of the proposed method and report the mean accuracy}. The recordings contain single improvised or scripted sentences uttered by each actor. Each utterance is annotated by 3 different people { and categorized into four emotional categories of \textit{Sad}, \textit{Happy}, \textit{Angry}, and \textit{Neutral}.} The length of the recordings are not standard throughout the dataset. Therefor we fix the length of the recordings to 4 seconds by cutting the longer utterances and padding the shorter utterances by repetition.

\subsection{Training Details}
The teacher and student networks are trained for 50 epochs on the same dataset. For the optimizer we use Adam optimizer \cite{kingma2015adam} with $\beta1=0.9$ and $\beta2=0.99$. We use cyclical learning rates \cite{smith2017cyclical} to train the networks with the initial learning rate of $10^{-4}$. We choose cyclical learning rates in order to decrease the probability of getting trapped in local minima. All networks are trained on a single Nvidia Titan RTX (24 GB vRAM) GPU with batch size of 32.

\subsection{Baselines} 

 {We compare our method with two re-implemented baselines: 
(1) ``Soft-label distillation'', which uses the soft-labels generated by the teacher network from video modality to train the student network. This approach has been used in knowledge distillation studies such as \cite{albanie2018emotion, nagrani2020disentangled, Nawaz_2021_CVPR};  In this baseline the parameter $\alpha$ (see Equation \ref{eq:normal_lupi}) determines how much the student model should follow the teacher.
(2) ``Multitask Learning'', which we described in Section \ref{sec:relate_lupi_wo}. This approach has been previously used in \cite{shi2017learning} for LUPI to perform action recognition from videos. In this case, the student model is trained using the ground-truth labels and gradients returning from a secondary decoder component which is tasked with generating a representation of the training sample in the feature space of the secondary modality. Here the parameter $\alpha$ represents the weight that is put on the gradients coming from the decoder component. }

\begin{table*}[!t]
\centering
\footnotesize
\caption{Experiment results for speaker identification on the Video version of VoxCeleb. Here, ($Acc_T$): Accuracy of the teacher network; ($Acc_S$): Accuracy of the student network; ($\Delta Acc_S$): The difference in the accuracy of the student network after distillation; {EM: The embedding-level implementation of the proposed model; AGG: The aggregation-level implementation of the proposed model; The values for proposed models are reported with  $\alpha=0.5$.}}
\label{tab:VoxCeleb_results}
\begin{tabular}{l|l|l|l|l} 
\hline
Model                                               & $Acc_T$          & \multiline{Acc_S}{w/o\, Distill.}      & \multiline{Acc_S}{w/\, Distill.}        & $\Delta Acc_S$\% \\ \hline \hline
Proposed (VGG)                                                    & 78.4               & 75.5                 		         & 77.5                                   & 2.64             \\ 
Proposed (VGG+BiLSTM+EM)                                          & 80.4               & 73.7                 		         & 75.3                                   & 2.17             \\ 
\textbf{Proposed (VGG+BiLSTM+AGG)}                                & 80.4               & 73.7                 		         & 76.4                                   & \textbf{3.66}             \\
Proposed (ResNet)                                                 & 83.1               & 80.3                 		         & 81.5                                   & 1.49             \\ 
Proposed (ResNet+BiLSTM+EM)                                       & 83.6               & 80.9                 		         & 82.2                                   & 1.60             \\ 
Proposed (ResNet+BiLSTM+AGG)                                      & 83.6               & 80.5                 		         & 82.7          	                      & 2.73         	\\
Proposed (SEResNet)                                               & 84.0               & 81.5                 		         & 82.2                                   & 0.85             \\ 
Proposed (SEResNet+BiLSTM+EM)                                     & 84.4               & 80.3                 		         & 80.8                                   & 0.62             \\ 
Proposed (SEResNet+BiLSTM+AGG)                                    & 84.4               & 80.5                 		         & 82.7          	                      & 2.73         	\\ \hline

\end{tabular}
\end{table*}

\begin{table*}[!t]
\centering
\footnotesize
\caption{Experiment results for speaker verification on VoxCeleb1 standard test set. Here, ($EER_S$): The equal error rate of the student network; ($\Delta EER_S$): The difference in the equal error rate of the student network after distillation. {EM: The embedding-level implementation of the proposed model; AGG: The aggregation-level implementation of the proposed model; The values for proposed models are reported with  $\alpha=0.5$.}}
\label{tab:VoxCeleb_results_2}
\begin{tabular}{l|l|l|l} 
\hline
Model                                                         & \multiline{EER_S}{w/o\, Distill.}    & \multiline{EER_S}{w/\, Distill.}       & $\Delta EER_S$\% \\ \hline \hline
Proposed (VGG)                                                & 10.15                 		         & 9.7                                    & 4.43             \\ 
Proposed (VGG+BiLSTM+EM)                                      & 9.34                 		         & 8.12                                   & 13.06             \\ 
Proposed (VGG+BiLSTM+AGG)                            & 9.34                 		         & 8.06                                   & 13.70             \\
Proposed (ResNet)                                             & 5.33                 		         & 3.91                                   & \textbf{26.64}             \\ 
Proposed (ResNet+BiLSTM+EM)                                   & 5.36                 		         & 4.05                                   & 24.44             \\ 
Proposed (ResNet+BiLSTM+AGG)                                  & 5.36                 		         & 3.95          	                      & 26.30         	\\
Proposed (SEResNet)                                           & 5.31                 		         & 4.22                                   & 20.52             \\ 
Proposed (SEResNet+BiLSTM+EM)                                 & 5.41                 		         & 4.35                                   & 19.59             \\ 
Proposed (SEResNet+BiLSTM+AGG)                                & 5.41                 		         & 4.27          	                      & 21.07         	\\ \hline

\end{tabular}
\end{table*}

\subsection{Performance}

\noindent\textbf{Speaker Recognition.} We use identification accuracy and equal error rate to evaluate the performance of the student model when the privileged information has been integrated into the framework.

We exclude other works such as \cite{lambert2018deep} for the reason that their proposed method requires the privileged information and original training data to be from the same modality. {\mbox{Figure \ref{fig:alpha_speaker}} shows the results of our experiments on the performance student models after using the proposed method and compares it with the baseline methods for different values of $\alpha$ (imitation parameter) and different model architectures.}  The figure shows that our proposed method has a positive impact on the performance of the student network for all values of $\alpha$. Moreover, we observe that the baseline methods show a negative impact when $\alpha$ is increased. This shows that while the baseline models exhibit successful performances when video is used to provide supervision to the audio in the absence of ground-truth labels \cite{albanie2018emotion, nagrani2020disentangled, Nawaz_2021_CVPR}, they do not provide any benefits for the scenario where supervised training is performed with both modalities but inference is performed only on audio.

It can also be observed that our method has the highest impact when $\alpha$ is either $0.5$ or $0.6$. This indicates that the best performance gain is achieved when the weight of the privileged information distillation on training of the networks is almost equal to that of the ground-truth labels.

\begin{figure*}[!t]
  \centering
  \includegraphics[width=.9\textwidth]{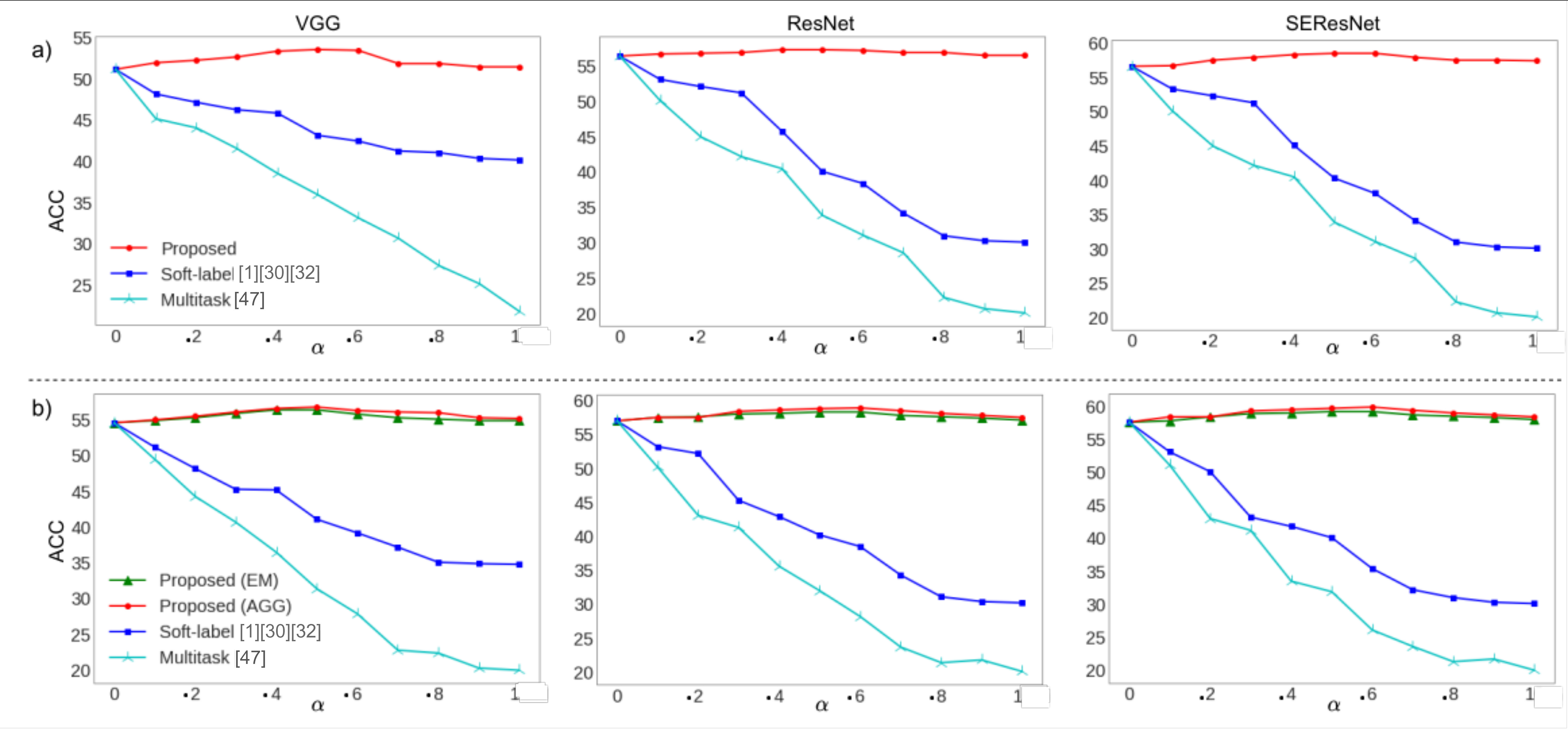}
  \caption{{A comparison between the effect of using our proposed method for LUPI versus using Soft-label and multitask training for SER on RAVDESS in relation to different values of $\alpha$ in: (a) Non-Sequential settings; (b) Sequential settings.}}
  \label{fig:alpha_ravdess}
\end{figure*}

\begin{table*}[!ht]
\centering
\footnotesize
\caption{Experiment results for SER (RAVDESS). Here, ($Acc_T$): Accuracy of the teacher network; ($Acc_S$): Accuracy of the student network; ($Acc_T - Acc_S$): The difference between the accuracy of the teacher and student network before distillation; ($\Delta Acc_S$): The difference in the accuracy of the student network after distillation; {EM: The embedding-level implementation of the proposed model; AGG: The aggregation-level implementation of the proposed model; The values for proposed models are reported with  $\alpha=0.5$.}}
\begin{tabular}{l|l|l|l|l} 
\hline
Model                                                                   & $Acc_T$           & \multiline{Acc_S}{w/o\, Distill.}      & \multiline{Acc_S}{w\, Distill.}      & $\Delta Acc_S$\% \\ \hline \hline
Proposed (VGG)                                   					    & 80.5              & 51.0                        & 52.3                                              & 2.54  \\ 
Proposed (VGG+BiLSTM+EM)                         					    & 83.4              & 54.6                        & 56.4                                              & 3.29  \\ 
Proposed (VGG+BiLSTM+AGG)                        					    & 83.4              & 54.6                        & 56.8                                              & 4.02  \\ 
Proposed (ResNet)                                					    & 81.1              & 56.4                        & 57.3                                              & 1.59  \\ 
Proposed (ResNet+BiLSTM+EM)                      					    & 84.6              & 56.9                        & 58.2                                              & 2.28  \\ 
Proposed (ResNet+BiLSTM+AGG)                     					    & 84.6              & 56.9                        & 58.8                                              & 3.33  \\ 
\textbf{Proposed (SEResNet)}                              			    & 81.0              & 56.6                        & 59.5                                              & \textbf{5.12}  \\ 
Proposed (SEResNet+BiLSTM+EM)                    					    & 85.3              & 57.6                        & 59.2                                              & 2.77  \\ 
Proposed (SEResNet+BiLSTM+AGG)                   				        & 85.3              & 57.6                        & 59.9                                              & 3.45  \\ \hline

\end{tabular}
\label{tab:RAVDESS}
\end{table*}

Table \ref{tab:VoxCeleb_results} shows the result of our experiment for integration of our method in different architectures for speaker identification on the video version of VoxCeleb dataset. We present the accuracy of the teacher $Acc_{T}$ for person identification task using the video modality. This will allow us to investigate the impact of our proposed method on the student for different teachers with varying performances. We also show the accuracy of the student network without distillation of privileged information so that we can better observe the performance gain using the proposed method and compare it to other methods. Lastly, we show the accuracy of the student network after distillation of privileged information and compare it with the student model without distillation ($\Delta Acc_S$). As shown by the results, we observe a substantial performance increase in student networks while using the proposed method. The highest performance gain is obtained when using the sequential networks with VGG-based encoders and distillation of privileged information on the aggregator. This is achieved while the accuracy of the teacher compared to the accuracy of the student is at its highest value, i.e., 9.09\%. We also observe that the highest performance gain, in comparison with the difference in the performance of teacher and student networks, occurs with the sequential network with a ResNet-based encoder and privileged information distillation on the aggregator. 

{Table \ref{tab:VoxCeleb_results_2} presents the results of our experiments for the benchmark networks for speaker verification {on the VoxCeleb1\cite{nagrani2020voxceleb} standard test set which includes recordings from 40 speakers outside of the training set}. We present the error rate of the student model by $EER_S$, before and after distillation, and compare the performance of the student model at these two points by calculating the difference and normalizing it by the EER w/o distillation ($\Delta EER_S\%$). We observe that the highest decrease in the error rate has been obtained by the ResNet-based encoders when the non-sequential implementation is used, indicating that these encoders often benefit more from being trained by the teacher models compared to other networks. }

\noindent\textbf{Emotion Recognition.} 
We use unweighted accuracy as the metric to evaluate the performance of the networks trained using our method for SER on RAVDESS. We compare our method to the re-implemented baselines described earlier for all the values of $\alpha$. \mbox{Figure \ref{fig:alpha_ravdess}} shows the results of this experiment. We observe that while our method exhibits a positive impact on the performance of the student networks, the baseline methods have a negative impact. This further shows that while the baseline models that utilize the video as the only source of supervision for learning audio are successful when the ground-truth labels are not present, they do not improve the performance of deep neural networks when training is done using both modalities but only audio is available at inference.

We also extend out experiments by comparing the performance of the proposed method integrated into different architectures. Table \ref{tab:RAVDESS} shows the result of this evaluation. We include the accuracy of the teacher network along with the accuracy of the student before and after distillation. It can be observed that using our method, the performance of the student models improve and the highest increase in the performance is achieved when the non-sequential student network using a SEResNet-based architecture is employed. We can also observe a similar behaviour to that of the previous experiment when comparing our method with the baselines. 

\begin{figure*}[!t]
  \centering
  \includegraphics[width=.9\textwidth]{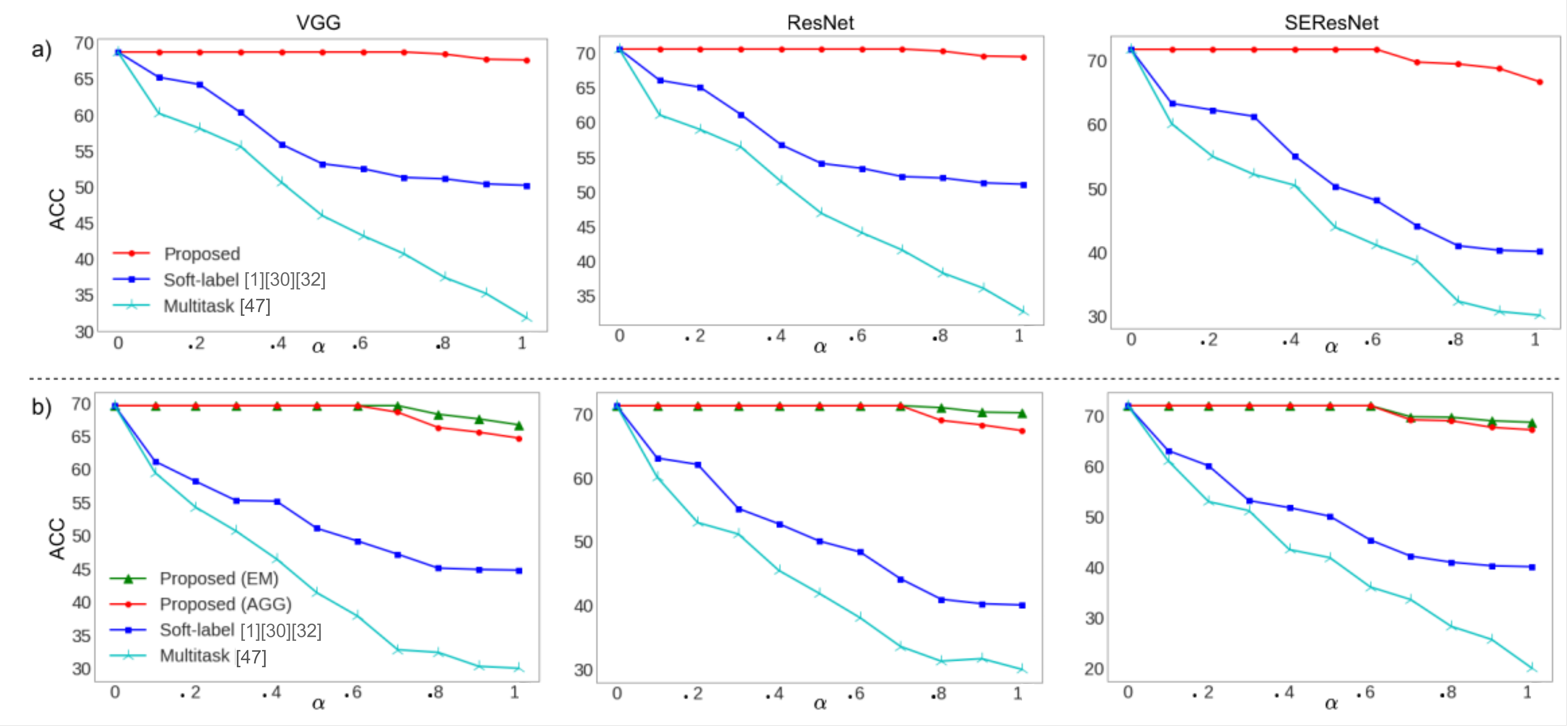}
  \caption{{A comparison between the effect of using our proposed method for LUPI versus using Soft-label and multitask training for SER on IEMOCAP in relation to different values of $\alpha$ in: (a) Non-Sequential settings; (b) Sequential settings.}}
  \label{fig:alpha_iemocap}
\end{figure*}

\begin{table*}[!t]
\centering
\footnotesize
\caption{Experiment results for SER (IEMOCAP). Here, ($Acc_T$): Accuracy of the teacher network; ($Acc_S$): Accuracy of the student network; ($Acc_T - Acc_S$): The difference between the accuracy of the teacher and student network before distillation; ($\Delta Acc_S$): The difference in the accuracy of the student network after distillation; {EM: The embedding-level implementation of the proposed model; AGG: The aggregation-level implementation of the proposed model; The values for proposed models are reported with  $\alpha=0.5$.}}
\begin{tabular}{l|l|l|l|l} 
\hline
Model                                                    & $Acc_T$          & \multiline{Acc_S}{w/o\, Distill.}      & \multiline{Acc_S}{w\, Distill.}    & $\Delta Acc_S$\% \\ \hline \hline
Proposed (VGG)                                           & 60.8             & 68.6               & -25.65  &  0  \\ 
Proposed (VGG+BiLSTM+EM)                                 & 64.3             & 69.6               & -21.55  &  0  \\ 
Proposed (VGG+BiLSTM+AGG)                                & 64.3             & 69.6               & -19.39  &  0  \\ 
Proposed (ResNet)                                        & 62.3             & 70.2               & -17.94  &  0  \\ 
Proposed (ResNet+BiLSTM+EM)                              & 65.2             & 70.9               & -20.45  &  0  \\ 
Proposed (ResNet+BiLSTM+AGG)                             & 65.2             & 70.9               & -18.33  &  0  \\ 
Proposed (SEResNet)                                      & 62.0             & 71.3               & -28.47  &  0  \\ 
Proposed (SEResNet+BiLSTM+EM)                            & 65.5             & 71.8               & -23.95  &  0  \\ 
Proposed (SEResNet+BiLSTM+AGG)                           & 65.5             & 71.8               & -21.86  & 0 \\ \hline

\end{tabular}
\label{tab:IEMOCAP_results}
\end{table*}

Lastly we evaluate our method on the IEMOCAP dataset. In this experiment we intend to show the effect of our method {in cases where the accuracy of the teacher network is \textit{lower} than the student network.} We compare the proposed method to the baseline methods described earlier for all the values of $\alpha$. \mbox{Figure \ref{fig:alpha_iemocap}} shows the results of this experiment. As shown in the figure, when using the proposed method, the performance of the student networks are not negatively effected by weaker teacher networks for low values of $\alpha$, whereas in the baseline methods, the negative impact is observed from very early values of $\alpha$. Table \ref{tab:IEMOCAP_results} shows the performance of our method integrated into different architectures. We observe that the best performance is achieved using the SEResNet architecture equipped with BiLSTM layers, and despite the teacher having a lower performance than the student, the proposed method does not negatively affect the performance of this network.

\section{Conclusion and Future Work}
We use teacher-student knowledge distillation for LUPI in order to take advantage of both audio and video inputs for training deep neural networks, while only using audio at inference. 
In our framework, embeddings are first extracted from the video input using a teacher model. The embeddings alongside the ground-truth labels are then used to train the student. 
We integrate our method in two different settings for non-sequential and sequential data. In the non-sequential setting, both the teacher and student networks are constructed using and encoder and a task header. We use the embeddings generated by the encoder of the teacher to train the encoder of the student, as the task header of the student is trained using the ground-truth labels. In the sequential setting, an additional aggregation component is introduced to the teacher and student networks, which is placed between the encoder and task header. We use two sets of embeddings produced by the encoder and aggregation component of the teacher to train the encoder and aggregation component of the student respectively. Similar to the non-sequential setting, the task header of the student is trained using ground-truth labels. 
By performing experiments on two tasks of SR and SER we show that our proposed framework leads to considerable performance gains in the student compared to previous studies. {While the benchmark models rely on different aspects of the input for SR and SER, and thus different architectures exhibit different performances for each task, our method consistently improves the performance of the benchmarks}. In summary, our approach opens a new path towards integration of LUPI by means of knowledge distillation into deep {audio} representation learning using audio-visual data, when only audio is available at inference. 

Our work also introduces a new set of challenges for future work. An immediate step for future work would be to study the use of generative models such as Generative Adversarial Networks and Variational Autoencoders for creating the embeddings from the teacher model with better domain adaptation and generalization. More recent and upcoming approaches such as normalizing flows can also be explored in this context. 

\bibliographystyle{abbrv}
\bibliography{
                references/References
                }

\begin{thebibliography}{10}

\bibitem{albanie2018emotion}
S.~Albanie, A.~Nagrani, A.~Vedaldi, and A.~Zisserman.
\newblock Emotion recognition in speech using cross-modal transfer in the wild.
\newblock {\em The ACM International Conference on Multimedia}, pages 292--301,
  2018.

\bibitem{atmaja2020multitask}
B.~T. Atmaja and M.~Akagi.
\newblock Multitask learning and multistage fusion for dimensional audiovisual
  emotion recognition.
\newblock {\em IEEE International Conference on Acoustics, Speech and Signal
  Processing (ICASSP)}, pages 4482--4486, 2020.

\bibitem{busso2008iemocap}
C.~Busso, M.~Bulut, C.-C. Lee, A.~Kazemzadeh, E.~Mower, S.~Kim, J.~N. Chang,
  S.~Lee, and S.~S. Narayanan.
\newblock Iemocap: Interactive emotional dyadic motion capture database.
\newblock {\em Language Resources and Evaluation}, 42(4):335, 2008.

\bibitem{chen2021cross}
D.~Chen, J.-P. Mei, Y.~Zhang, C.~Wang, Z.~Wang, Y.~Feng, and C.~Chen.
\newblock Cross-layer distillation with semantic calibration.
\newblock {\em The AAAI Conference on Artificial Intelligence},
  35(8):7028--7036, 2021.

\bibitem{do2019compact}
T.~Do, T.-T. Do, H.~Tran, E.~Tjiputra, and Q.~D. Tran.
\newblock Compact trilinear interaction for visual question answering.
\newblock {\em IEEE/CVF International Conference on Computer Vision (ICCV)},
  pages 392--401, 2019.

\bibitem{gao2021residual}
M.~Gao, Y.~Wang, and L.~Wan.
\newblock Residual error based knowledge distillation.
\newblock {\em Neurocomputing}, 433:154--161, 2021.

\bibitem{garcia2018modality}
N.~C. Garcia, P.~Morerio, and V.~Murino.
\newblock Modality distillation with multiple stream networks for action
  recognition.
\newblock {\em The European Conference on Computer Vision (ECCV)}, pages
  103--118, 2018.

\bibitem{guan2020differentiable}
Y.~Guan, P.~Zhao, B.~Wang, Y.~Zhang, C.~Yao, K.~Bian, and J.~Tang.
\newblock Differentiable feature aggregation search for knowledge distillation.
\newblock {\em European Conference on Computer Vision (ECCV)}, Part XXIV
  16:469--484, 2020.

\bibitem{CNN_speech_mine_short}
A.~Hajavi and A.~Etemad.
\newblock A deep neural network for short-segment speaker recognition.
\newblock {\em {INTERSPEECH}}, pages 2878--2882, 2019.

\bibitem{CNN_speech_mine}
A.~Hajavi and A.~Etemad.
\newblock Knowing what to listen to: Early attention for deep speech
  representation learning.
\newblock {\em arXiv preprint arXiv:2009.01822}, 2020.

\bibitem{hajavi2021siamese}
A.~Hajavi and A.~Etemad.
\newblock Siamese capsule network for end-to-end speaker recognition in the
  wild.
\newblock {\em IEEE International Conference on Acoustics, Speech and Signal
  Processing (ICASSP)}, pages 7203--7207, 2021.

\bibitem{he_deep_2016}
K.~He, X.~Zhang, S.~Ren, and J.~Sun.
\newblock {Deep Residual Learning for Image Recognition}.
\newblock {\em The IEEE/CVF Conference on Computer Vision and Pattern
  Recognition (CVPR)}, pages 770--778, 2016.

\bibitem{heo2019comprehensive}
B.~Heo, J.~Kim, S.~Yun, H.~Park, N.~Kwak, and J.~Y. Choi.
\newblock A comprehensive overhaul of feature distillation.
\newblock {\em The IEEE/CVF International Conference on Computer Vision
  (ICCV)}, pages 1921--1930, 2019.

\bibitem{heo2019knowledge}
B.~Heo, M.~Lee, S.~Yun, and J.~Y. Choi.
\newblock Knowledge transfer via distillation of activation boundaries formed
  by hidden neurons.
\newblock {\em The AAAI Conference on Artificial Intelligence},
  33(01):3779--3787, 2019.

\bibitem{hinton2015distilling}
G.~Hinton, O.~Vinyals, J.~Dean, et~al.
\newblock Distilling the knowledge in a neural network.
\newblock {\em arXiv preprint arXiv:1503.02531}, 2(7), 2015.

\bibitem{hu2018squeeze}
J.~Hu, L.~Shen, and G.~Sun.
\newblock Squeeze-and-excitation networks.
\newblock {\em The IEEE/CVF Conference on Computer Vision and Pattern
  Recognition (CVPR)}, pages 7132--7141, 2018.

\bibitem{huang2017like}
Z.~Huang and N.~Wang.
\newblock Like what you like: Knowledge distill via neuron selectivity
  transfer.
\newblock {\em arXiv preprint arXiv:1707.01219}, 2017.

\bibitem{jalal2019learning}
M.~A. Jalal, E.~Loweimi, R.~K. Moore, and T.~Hain.
\newblock Learning temporal clusters using capsule routing for speech emotion
  recognition.
\newblock {\em {INTERSPEECH}}, pages 1701--1705, 2019.

\bibitem{kansizoglou2019active}
I.~Kansizoglou, L.~Bampis, and A.~Gasteratos.
\newblock An active learning paradigm for online audio-visual emotion
  recognition.
\newblock {\em IEEE Transactions on Affective Computing}, 1:1, 2019.

\bibitem{kim2018paraphrasing}
J.~Kim, S.~Park, and N.~Kwak.
\newblock Paraphrasing complex network: network compression via factor
  transfer.
\newblock {\em International Conference on Neural Information Processing
  Systems (NeurIPS)}, pages 2765--2774, 2018.

\bibitem{kingma2015adam}
D.~P. Kingma and J.~Ba.
\newblock Adam: A method for stochastic optimization.
\newblock {\em International Conference on Learning Representations (ICLR)},
  2015.

\bibitem{komodakis2017paying}
N.~Komodakis and S.~Zagoruyko.
\newblock Paying more attention to attention: improving the performance of
  convolutional neural networks via attention transfer.
\newblock {\em International Conference on Learning Representations (ICLR)},
  2017.

\bibitem{tedd1}
T.~Kourkounakis, A.~Hajavi, and A.~Etemad.
\newblock Detecting multiple speech disfluencies using a deep residual network
  with bidirectional long short-term memory.
\newblock {\em IEEE International Conference on Acoustics, Speech and Signal
  Processing (ICASSP)}, pages 6089--6093, 2020.

\bibitem{tedd2}
T.~Kourkounakis, A.~Hajavi, and A.~Etemad.
\newblock Fluentnet: End-to-end detection of stuttered speech disfluencies with
  deep learning.
\newblock {\em IEEE/ACM Transactions on Audio, Speech, and Language
  Processing}, 29:2986--2999, 2021.

\bibitem{kumar2021towards}
P.~Kumar, V.~Kaushik, and B.~Raman.
\newblock Towards the explainability of multimodal speech emotion recognition.
\newblock {\em {INTERSPEECH}}, pages 1748--1752, 2021.

\bibitem{lambert2018deep}
J.~Lambert, O.~Sener, and S.~Savarese.
\newblock Deep learning under privileged information using heteroscedastic
  dropout.
\newblock {\em IEEE Conference on Computer Vision and Pattern Recognition},
  pages 8886--8895, 2018.

\bibitem{livingstone2018ryerson}
S.~R. Livingstone and F.~A. Russo.
\newblock The ryerson audio-visual database of emotional speech and song
  (ravdess): A dynamic, multimodal set of facial and vocal expressions in north
  american english.
\newblock {\em PloS One}, 13(5):e0196391, 2018.

\bibitem{vapnik2016LUPIMAIN}
D.~Lopez-Paz, L.~Bottou, B.~Schölkopf, and V.~Vapnik.
\newblock Unifying distillation and privileged information.
\newblock {\em International Conference on Learning Representations (ICLR)},
  pages 1--10, 2016.

\bibitem{Nagrani_2018_ECCV}
A.~Nagrani, S.~Albanie, and A.~Zisserman.
\newblock Learnable pins: Cross-modal embeddings for person identity.
\newblock {\em The European Conference on Computer Vision (ECCV)}, 2018.

\bibitem{nagrani2020disentangled}
A.~Nagrani, J.~S. Chung, S.~Albanie, and A.~Zisserman.
\newblock Disentangled speech embeddings using cross-modal self-supervision.
\newblock {\em IEEE International Conference on Acoustics, Speech and Signal
  Processing (ICASSP)}, pages 6829--6833, 2020.

\bibitem{nagrani2020voxceleb}
A.~Nagrani, J.~S. Chung, W.~Xie, and A.~Zisserman.
\newblock Voxceleb: Large-scale speaker verification in the wild.
\newblock {\em Computer Speech \& Language}, 60:101027, 2020.

\bibitem{Nawaz_2021_CVPR}
S.~Nawaz, M.~S. Saeed, P.~Morerio, A.~Mahmood, I.~Gallo, M.~H. Yousaf, and
  A.~Del~Bue.
\newblock Cross-modal speaker verification and recognition: A multilingual
  perspective.
\newblock {\em The IEEE/CVF Conference on Computer Vision and Pattern
  Recognition (CVPR) Workshops}, pages 1682--1691, June 2021.

\bibitem{passalis2018learning}
N.~Passalis and A.~Tefas.
\newblock Learning deep representations with probabilistic knowledge transfer.
\newblock {\em The European Conference on Computer Vision (ECCV)}, pages
  268--284, 2018.

\bibitem{passban2021alp}
P.~Passban, Y.~Wu, M.~Rezagholizadeh, and Q.~Liu.
\newblock Alp-kd: Attention-based layer projection for knowledge distillation.
\newblock {\em The AAAI Conference on Artificial Intelligence},
  35(15):13657--13665, 2021.

\bibitem{roheda2018cross}
S.~Roheda, B.~S. Riggan, H.~Krim, and L.~Dai.
\newblock Cross-modality distillation: A case for conditional generative
  adversarial networks.
\newblock {\em IEEE International Conference on Acoustics, Speech and Signal
  Processing (ICASSP)}, pages 2926--2930, 2018.

\bibitem{romero2015fitnets}
A.~Romero, N.~Ballas, S.~E. Kahou, A.~Chassang, C.~Gatta, and Y.~Bengio.
\newblock Fitnets: Hints for thin deep nets.
\newblock {\em International Conference on Learning Representations (ICLR)},
  2015.

\bibitem{shi2017learning}
Z.~Shi and T.-K. Kim.
\newblock Learning and refining of privileged information-based rnns for action
  recognition from depth sequences.
\newblock {\em The IEEE/CVF Conference on Computer Vision and Pattern
  Recognition (CVPR)}, pages 3461--3470, 2017.

\bibitem{simonyan2014very}
K.~Simonyan and A.~Zisserman.
\newblock Very deep convolutional networks for large-scale image recognition.
\newblock {\em International Conference on Learning Representations (ICLR)},
  2015.

\bibitem{smith2017cyclical}
L.~N. Smith.
\newblock Cyclical learning rates for training neural networks.
\newblock {\em IEEE Winter Conference on Applications of Computer Vision
  (WACV)}, pages 464--472, 2017.

\bibitem{thoker2019cross}
F.~M. Thoker and J.~Gall.
\newblock Cross-modal knowledge distillation for action recognition.
\newblock {\em IEEE International Conference on Image Processing (ICIP)}, pages
  6--10, 2019.

\bibitem{tian2019contrastive}
Y.~Tian, D.~Krishnan, and P.~Isola.
\newblock Contrastive representation distillation.
\newblock {\em arXiv preprint arXiv:1910.10699}, 2019.

\bibitem{vapnik2009LUPI1}
V.~Vapnik and A.~Vashist.
\newblock A new learning paradigm: Learning using privileged information.
\newblock {\em Neural Networks}, 22(5-6):544--557, 2009.

\bibitem{xie2019utterance}
W.~Xie, A.~Nagrani, J.~S. Chung, and A.~Zisserman.
\newblock {Utterance-level Aggregation For Speaker Recognition In The Wild}.
\newblock {\em IEEE International Conference on Acoustics, Speech and Signal
  Processing (ICASSP)}, pages 5791--5795, 2019.

\bibitem{xu2020feature}
K.~Xu, L.~Rui, Y.~Li, and L.~Gu.
\newblock Feature normalized knowledge distillation for image classification.
\newblock {\em European Conference on Computer Vision (ECCV)}, Part XXIV
  16:664--680, 2020.

\bibitem{you2017learning}
S.~You, C.~Xu, C.~Xu, and D.~Tao.
\newblock Learning from multiple teacher networks.
\newblock {\em The ACM International Conference on Knowledge Discovery and Data
  Mining (SIGKDD)}, pages 1285--1294, 2017.

\bibitem{zhang2018better}
C.~Zhang and Y.~Peng.
\newblock Better and faster: knowledge transfer from multiple self-supervised
  learning tasks via graph distillation for video classification.
\newblock {\em The International Joint Conference on Artificial Intelligence
  (IJCAI)}, pages 1135--1141, 2018.

\bibitem{zhou2018rocket}
G.~Zhou, Y.~Fan, R.~Cui, W.~Bian, X.~Zhu, and K.~Gai.
\newblock Rocket launching: A universal and efficient framework for training
  well-performing light net.
\newblock {\em The AAAI Conference on Artificial Intelligence}, 2018.

\end{thebibliography}

\end{document}